\documentclass{ws-procs9x6}

\def\etal {{\it et al.}}

\begin{document}

\title{SPONTANEOUS LORENTZ SYMMETRY BREAKING\\
IN NONLINEAR ELECTRODYNAMICS}
\author{L.F.\ URRUTIA }
\address{Instituto de Ciencias Nucleares, Universidad Nacional Aut\'onoma de M\'exico\\
Ciudad de M\'exico, 04510 D.F., M\'exico\\
E-mail: urrutia@nucleares.unam.mx}
\begin{abstract}
We review some of the basic features and predictions of a gauge invariant spontaneous Lorentz
symmetry breaking model arising from the nonzero vacuum expectation value of the electromagnetic
tensor and leading to a nonlinear electrodynamics. The model is stable in the small Lorentz
invariance violation approximation. The speed of light is independent of the frequency
and one of the propagating modes is highly anisotropic. The bound $\Delta c/c \, < \, 10^{-32}$
is obtained for such anisotropy measured in perpendicular directions.
\end{abstract}
\bodymatter
\section{Introduction}
Many candidate theories for describing the structure  of spacetime
at the microscopic level, like string theory, models of quantum
gravity and noncommutative theories, for example, lead to the
picture that spacetime has a discrete nature for very small scales,
instead of the continuum description in which most modern physics is
based. This poses the  natural question of whether or not such
granular structure will leave measurable imprints upon the dynamics
of particles at Standard Model energies. The analogy of particle
propagation in crystals suggests that modifications will indeed
arise. Thus, one of the open problems of these spacetime theories is
to determine the nature of these modifications, in case they are
produced. The possibility that such corrections may incorporate
Lorentz invariance violation (LIV)  was suggested in Ref.
\refcite{GAC} and it has recently been the subject of intense study
through astrophysical observations \cite{OBSTD}. Moreover, some
heuristic calculations \cite{HLQG}, inspired in loop quantum
gravity, provide also support to this conjecture. This possibility
adds additional interest to the search for LIV, specially given that
many observations and experiments have already attained Planck scale
sensitivities. In such a way, these results will serve as physical
constraints  to select the right quantum theory of spacetime among
the competing proposals, once they are able to fill the gap between
the quantum and the semiclassical scales. Since this goal has not
yet been achieved and even though the contact between these two
regimes may require the introduction of a completely different
conceptual structure in modern physics, standard effective quantum
field theories, in the form of the so called Standard-Model
Extension (SME) \cite{SME}, provide an adequate tool to study  such
modifications at Standard Model  energies.

In this contribution, the work done in collaboration with J. Alfaro
\cite{JALU} regarding a model of nonlinear gauge invariant
electrodynamics arising from spontaneous Lorentz symmetry breaking
(SLSB) is reviewed. This model is complementary to related  studies
of SLSB in the literature. On one hand there are  theories where the
photon emerges as the Goldstone boson of such breaking and which
allow to recover electrodynamics, in a nonlinear gauge, at the tree
and one loop level, thus providing a dynamical setting for $U(1)$
gauge invariance \cite{PHOTONSGB}. Also, models with SLSB arising
from the vacuum expectation value (VEV) of  antisymmetric tensors
$B_{\mu\nu}$ coupled to gravity have been studied \cite{BAQBAK}. Here
the two-form field $B=B_{\mu\nu}dx^\mu \wedge dx^\nu $ is considered
as the potential producing the  field strength $H=dB$ that enters in
the kinetic term of the action. The model  described in this
contribution is  intermediate among those two: gauge invariance is
always preserved, SLSB is induced by a VEV of $F_{\mu\nu}$ whose
excitations around the minimum turn out to be the usual
electromagnetic field, which is ultimately described by a vector
potential $A_\mu$ with the standard kinetic term for
electrodynamics. The interpretation of the Goldstone mechanism in
our case differs from the standard one related to massless
excitations and its description is postponed for future work.
\section{The model }
\label{aba:sec1}
We start from the Lagrangian
\begin{equation}
L(F_{\alpha \beta },X_{\mu })=-V(F_{\alpha \beta })-\bar{F}^{\nu \mu
}\partial _{\nu }X_{\mu }, \quad F_{\alpha \beta }=-F_{\beta \alpha
},
\label{LAGINI}
\end{equation}%
where $\bar{F}_{\mu \nu}$ is the dual of ${F}_{\mu \nu}$. The fields
$X_\mu$ are Lagrange multipliers which ultimately will impose the
condition that the excitations of ${F}_{\mu \nu}$ are derived from a
vector potential, thus recovering a nonlinear electrodynamics. The
potential $V(F)$ provides a minimum for the VEV $C_{\mu\nu}$ of
${F}_{\mu \nu}$. In Ref. \refcite{JALU} we have made plausible
the appearance of such a potential, starting from a conventional
gauge theory including fermions, gauge fields and Higgs fields which
provide masses to the gauge bosons, except for the photon potential
${\tilde A}_{\mu }$. For our purposes here it is enough to start
with the  standard Ginzburg-Landau parametrization of such potential
\begin{equation}
V(F_{\mu \nu })=\frac{1}{2}\alpha F^{2}+\frac{\beta }{4}\left(
F^{2}\right) ^{2}, \quad \beta\, > \, 0.  \label{POT}
\end{equation}%
The vacuum configuration $C_{\mu\nu}, C_\mu$ is obtained by minimizing the energy
of the system, obtained from the Lagrangian (\ref{LAGINI}) via Noether's theorem,
and requiring constant field configurations in order to preserve translational
invariance. The action for the excitations $a_{\alpha\beta}$ and ${\bar X}_\mu$
around such minima is subsequently obtained and the elimination of the Lagrange
multiplier ${\bar X}_\mu$ introduces the potential $A_\mu$ such
that $a_{\mu\nu}=\partial_\mu A_\nu-\partial_\nu A_\mu + l C_{\mu\nu}$,
where $l$ is a constant. After making some rescaling we arrive at the action
\begin{equation}
S(A_\alpha)=\int d^{4}x\left({-\frac{\left[1-D^{2}{\cal B}\right]}{4} D^{2}}-%
\frac{f_{\mu \nu }f^{\mu \nu }}{4}-{\cal B}\left[\frac{}{}  D_{\mu \nu
}f^{\mu \nu } + f_{\mu \nu }f^{\mu \nu } \right]
^{2}\right),
\label{ACFIN}
\end{equation}
which defines the model.
Here $f_{\mu\nu}=\partial_\mu A_\nu-\partial_\nu A_\mu$; $D_{\mu\nu}$, which replaces
the VEV $C_{\mu\nu}$,  is the arbitrary  constant antisymmetric tensor characterizing
the vacuum, $D^2=D_{\mu\nu}D^{\mu\nu}$ and ${\cal B}$ is a positive constant.
\section{The symmetry algebras of the broken theory}
\label{sec3} As proposed in Ref. \refcite{BAQBAK}, the simplest
parametrization of the vacuum $D_{\mu\nu}$, written in terms of the
usual electric and magnetic components, is given by two independent
quantities for each of the following cases: (i)
$\mathbf{e}=\{0,0,e\}, \, \mathbf{b}=\{0,0,b\}$, when at least one
of the electromagnetic invariants is not zero (the choice $\psi=0$
in Ref. \refcite{JALU}) and (ii) $\mathbf{e}=\{0,e,0\}, \,
\mathbf{b}=\{0,0,b\}$, when both electromagnetic invariants are zero
(the choice $\psi=\pi/2$ in Ref. \refcite{JALU}). The remaining
symmetries of the broken action are obtained by requiring that the
vacuum be invariant under the transformations $
G_{\hspace{0.75em}\alpha }^{\mu }D^{\alpha \nu }+G_{\;\alpha }^{\nu
}\;D^{\mu \alpha },  \label{COND3} $ generated by $G_{\;\alpha
}^{\nu}$ which denote the standard infinitesimal Lorentz algebra
generators,  plus dilation transformations ($x^\mu \partial_\mu$)
which are represented by a multiple of the identity in this
restricted algebra. The case (i) leads to $T(2)$ as the remaining
symmetry algebra, while the case (ii) leads to  HOM(2).
\section{Dispersion relations and polarizations}
The propagation properties of the model arise from the quadratic terms in
the effective Lagrangian%
\begin{equation}
L_{0}=-\frac{1}{4}f_{\mu \nu }f^{\mu \nu }-{\cal B}\left( f_{\mu \nu }D^{\mu \nu
}\right) ^{2}.\;\;\;
\end{equation}%
The equations of motion are
\begin{equation}
\left( \partial ^{2}A_{\beta }-\partial _{\beta }\partial ^{\alpha
}A_{\alpha }\right) =-8{\cal B}D_{\alpha \beta }\partial ^{\alpha }\left( D^{\mu
\nu }\partial _{\mu }A_{\nu }\right).  \label{EQMOT}
\end{equation}%
Introducing the definitions
\begin{equation}
\mathbf{D}=\mathbf{E+}8{\cal B}\mathbf{e}(\ \mathbf{B\cdot b}-\mathbf{E\cdot e}%
),\;\;\;\;\mathbf{H}=\mathbf{B}+8{\cal B}\mathbf{b}(\mathbf{B\cdot b}-\mathbf{%
E\cdot e}),
\end{equation}
Eqs. (\ref{EQMOT}) adopt the standard form of Maxwell's equation in a medium. The dispersion
relations and polarization properties of a plane wave propagating with momentum $k_\alpha$
are: (1) when  {$D_{i\alpha}A^ik^\alpha=0$},
the triad {$ {\mathbf E},\ {\mathbf B},\ {\mathbf k}$} together with the dispersion
relation are  the standard ones; (2) when {$D_{i\alpha}A^ik^\alpha \neq 0$} we have
\begin{equation}
\mathbf{E} \cdot \mathbf{B}=0, \quad
\mathbf{k} \cdot {\mathbf B}=0,\quad
\mathbf{k} \cdot {\mathbf E}\neq 0 \quad \rm{and } \quad \omega=|\mathbf{k}|\times F(angles),
\end{equation}
where {angles} refer to those between $\mathbf{k}$ and the vectors characterizing the vacuum.
Here $A^i$ is in the Coulomb gauge.
In the approximation
${\cal B}{\mathbf e}^2, {\cal B}{\mathbf b}^2,  {\cal B}|{\mathbf e}||{\mathbf b}| \ll \, 1$,
the speed {$c_{1{\rm w}}(\mathbf{\hat{k}})=|\nabla_k \omega|$} in case (2) is
\begin{equation}
c_{1{\rm w}}(\mathbf{\hat{k}})=1+8{\cal B}\left( {\mathbf e}^{2}
+{\mathbf b}^{2}\right) -4{\cal B}\left( \left({%
\mathbf{b} \cdot \hat{\mathbf{k}}}\right) ^{2}+\left( {{\mathbf e} \cdot \hat{{\mathbf k}}}\right)^{2}-2\,
{\hat{{\mathbf k}}\, \cdot }\left({{\mathbf e}} \times \mathbf{b}\right) \right).
\end{equation}
\section{Embedding in the SME}
The propagating sector can be embedded in the SME via the identification
{\begin{equation}
-{\cal B}\left( f_{\mu \nu }D^{\mu \nu }\right) ^{2}=-\frac{1}{4}
\left(k_{F}\right)^{\kappa \lambda \mu \nu }f_{\kappa \lambda }f_{\mu \nu },
\end{equation}}
which produces
\begin{equation}
\left( k_{F}\right)^{\kappa \lambda \mu \nu }= 4{\cal B}D^{\kappa \lambda
}D^{\mu \nu }+ \left[ 2{\cal B} D^{\kappa \mu }D^{\lambda \nu }-\frac{1}{2}{\cal B}D^{2} \eta ^{\kappa
\mu }\eta ^{\lambda \nu }- ( \kappa \leftrightarrow \lambda)\right].
\label{KF}
\end{equation}
We have explicitly verified that the above realization of $\left(
k_{F}\right) ^{\kappa \lambda \mu \nu }$ satisfies all the required
identities. The relation (\ref{KF}) allows to express the components
of $\left( k_{F}\right) ^{\kappa \lambda \mu \nu }$ in terms of two
independent parameters ${\cal B}e^2, \, {\cal B}b^2 $ according to
the cases described at the beginning of Sec.\ \ref{sec3}. In this
way, the stringent astrophysical bounds \cite{KR} ${\bar
\kappa}^{ij}_{e+}, \, {\bar \kappa}^{ij}_{o-} < 10^{-32}$ are
summarized in the condition
\begin{equation}
{\cal B}\left( {\mathbf e}^{2}+{\mathbf b}^{2}\right) \, <  \, 2.5\times 10^{-33},
\end{equation}
which satisfies  all the less stringent remaining bounds. Defining  the two-way speed of light
$
c_{2{\rm w}}(\mathbf{\hat{k}})=[ c_{1{\rm w}}(\mathbf{\hat{k}})+c_{1{\rm w}}(-\mathbf{%
\hat{k}})]/2
$
leads to the following  bound upon the anisotropy of such velocity, measured along
perpendicular trajectories
\begin{equation}
{\Delta c}/{c} \equiv \left| c_{2{\rm w}}(\mathbf{\hat{k}})-c_{2{\rm w}}(\mathbf{%
\hat{q}})\right|/c\,\, < 10^{-32},  \quad    \hat{{\mathbf q}}=
\hat{{\mathbf k}}\times\left(\hat{{\mathbf k}}\times(\hat{{\mathbf e}}\times
\hat{{\mathbf b}} )\right).
\end{equation}
The recent bound $\left({\Delta c}/{c}\right)_{LAB} \sim 10^{-17}$
is  recovered provided we use the corresponding restrictions for the
accessible parameters in the laboratory ${\bar \kappa}^{ij}_{e-}, \,
{\bar \kappa}^{ij}_{o+}$ \cite{RECBOUND}. In Ref. \refcite{RALF} we
find the latest bound  for ${\bar \kappa}^{ij}_{o+}$ which is $1.6
\times 10^{-14}$. Assuming that the vacuum parameters $\mathbf{e}$,
$\mathbf{b}$ might represent some relic fields in the actual era,
and that the constant  $\rho \simeq {1}/{2}\left(
\mathbf{b}^{2}-\mathbf{e}^{2}\right)$ in  (\ref{ACFIN})  can be
associated with the cosmological constant $|\rho _{\Lambda
}|<\;10^{-48}\;\left( GeV\right) ^{4}$ \cite{CARROL}, we obtain the
bound $ |\mathbf{b|\;<\;}5\times 10^{-5}
$ Gauss,  by performing a passive Lorentz transformation to a reference frame where $%
\mathbf{e=0}$, which we assume to be concordant with the standard inertial reference
frame. This result is consistent with observations of intergalactic magnetic fields \cite{MF}.
\section*{Acknowledgments}
The author has been partially supported by the projects CONACyT \# 55310 and  UNAM-DGAPA \# IN111210.
He also thanks  V.A. Kosteleck{\'{y}} and collaborators for their splendid organization of CPT10.

\end{document}